# Experimental study on the cyclic resistance of a natural loess from Northern France

KARAM J.P.[1], CUI Y.J.[1], TANG A.M.[1], TERPEREAU J-M.[2], MARCHADIER G.[2*]

1 : Ecole des Ponts ParisTech, UR Navier/CERMES, France
2 : Société Nationale des Chemins de Fer (SNCF), France

**Corresponding author**
Prof. Yu Jun CUI
ENPC/CERMES
6 et 8, av. Blaise Pascal
Cité Descartes, Champs-sur-Marne
77455 Marne La Vallée cedex 2
France
Tel: 33 1 64 15 35 50
Fax: 33 1 64 15 35 62
E-mail: cui@cermes.enpc.fr

\* This work is dedicated to Gilles Marchadier who died on May 29th, 2007, at the age of 30.

*Abstract*

In order to analyze the instability phenomenon observed along the Northern High Speed Line of Réseau Ferré de France (RFF), soil blocks were taken at a site near the railway, at four different depths (1.2, 2.2, 3.5 and 4.9 m). Cyclic triaxial tests were carried out on saturated and unsaturated soil specimens. The results from tests on initially saturated specimens showed that the soil taken at 2.2 m depth has the lowest resistance to cyclic loading, in relation to its highest porosity and lowest clay fraction. This soil was then studied at unsaturated state with various initial water contents. Unsaturated soil specimens were first subjected to cyclic loadings to decrease their volume. These cyclic loadings was stopped when the volume decrease was approximately equal to the initial pore air volume, or when the pores filled by air were eliminated and the soil was considered to become saturated. Afterwards, the back-pressure tubing was saturated with de-aired water and cycles were applied under undrained condition. Significant effect of initial water content was evidenced: the lower the initial water content, the higher the cyclic resistance. This can be explained by the densification of the soil during the initial cyclic loadings.

**Key words**: *loess, cyclic resistance, liquefaction, instability, saturation, densification.*

**Introduction**

Northern French TGV (high speed train) railway is subjected to an equivalent 3500 cycles per day and around 1 270 000 cycles per year. Between 1999 and 2004, numerous sinkholes on ground surface after long rainy periods were observed along this line. These holes were about 5 m in diameter and 1 m deep. For the holes situated far from the railway, the origin was identified and attributed to the cavities that had existed since the First World War and other existing quarries. This represents 60% of the identified cases; the remaining 30% involved the zone at a short distance of 25 m away from the railway, for which the collapse could not be explained exclusively by the cavity presence. In 2003, numerical simulations by finite element modeling were performed by a company in order to assess the cyclic loads in the soil. It was reported that until 5 m away from the railway, the amplitude of vertical cyclic load varies from 11 to 23 kPa till 4 m depth at a frequency of 20 Hz. The load considerably decreases beyond 5 m.

From a mechanical point of view, the collapse of loess could be related to two deformation mechanisms: (i) collapse upon wetting; (ii) instability onset characterized by large deformation development when subjected to cyclic loading under saturated conditions. Obviously, these two mechanisms could be strongly affected by the fatigue of the soil due to large number of loading cycles applied by trains, in most cases under unsaturated conditions. Cui et al. (2004) studied the collapse mechanism by performing the simple and double oedometer tests (Jennings and Knight 1957, Knight 1963) and the results showed that the collapse susceptibility of the soil upon wetting was not high enough to explain the observed instability level in field. It is therefore necessary to explore the second mechanism. Indeed, under natural condition, it is possible that in a certain area the loessic soil becomes saturated, giving rise to instability onset because of their reduced mechanical resistance compared to their unsaturated state. It is also possible that the loessic soil, initially unsaturated, become saturated due to volume changes resulted from train-induced cyclic loadings.

Assessment of soil liquefaction and instability has been an important issue in the field of geotechnical engineering. Several approaches have been developed for liquefaction assessment. The most commonly used approaches are the stress (Seed and Idriss 1971) and the strain (Dobry et al. 1982) methods. The former aims at characterizing the earthquake loading and the soil resistance in terms of cyclic stresses, whereas in the latter, the characterization is performed in terms of cyclic strains. Cyclic triaxial tests, cyclic simple shear tests and cyclic torsion shear tests are used as common laboratory tests for soil characterization. As far as the liquefaction criteria are concerned, there are mainly two categories: i) pore water pressure-based criterion indicating that liquefaction occurs when zero effective net mean stress, p', is reached (Seed and Lee 1966), ii) strain/deformation-based criterion indicating that certain amount of single or double amplitude of axial strain, $\varepsilon_1$, is produced (Dobry et al. 1982; Ishihara 1993).

Castro (1975) and Castro and Poulos (1977) distinguished the liquefaction phenomenon from the cyclic mobility one, the former corresponding to the failure of saturated contractive sands subjected to cyclic shear stresses and the latter corresponding to the failure showed by saturated dilative sands. The two phenomena have been later on referred to as pre-failure instability. The instability is often defined as a condition that occurs when a soil element subjected to a small effective stress perturbation cannot sustain the current stress state imposed on it and a runaway deformation takes place (Lade and Pradel 1990, Vaid and Eliadorani 1998, Leong et al. 2000, Chu et al. 2003). In most cases, the instability phenomenon occurs when shearing the saturated sands under undrained conditions. However, it can also occur under drained conditions (Chu et al. 1993) or partially undrained conditions (Vaid and Eliadorani 1998). As Chu and Leong (2001) indicated, the instability phenomenon manifests only when shearing soil samples under a load-controlled mode, a deformation-controlled mode leading to strain softening. Moreover, the instability and strain softening are only different exhibitions of the same mechanism under different loading conditions.

Previous studies carried out in an attempt to understand the effect of partial saturation on the undrained cyclic shear strength of sand (Sherif et al. 1977; Chaney 1978; Yoshimi et al. 1989) have shown that soil cyclic resistance is strongly dependent on its initial saturation state: for a given cyclic stress ratio $R_{CR}$, the number of cycles causing liquefaction has been found to increase substantially with decreasing Skempton B coefficient. In earthquake conditions, Yang and Sato (2000), Yang and Sato (2001) observed that the partial saturation condition may give rise to greater amplification of vertical ground motion.

The present work aims at investing the instability mechanism by studying the cyclic behavior of the natural loess taken at a site close to the railways of northern TGV line. Emphasis was made on the effect of initial saturation state on the cyclic resistance by performing special tests in two stages: i) cyclic loading under air-drained and nearly constant water content condition to reach soil saturation, and ii) cyclic loading under undrained condition. Because no field recorded data have been available in terms of pre-loading history, as a first attempt, the confining pressure of 25 kPa was considered based on the calculation results mentioned previously. Note also that according to the field monitoring data, the cyclic deviator stress is estimated at 20 kPa. In order to explore the instability phenomenon of the soil taken from different depths or at different unsaturated conditions, higher cyclic deviator stresses were applied in order to drive the soil beyond the instability state. In this work, the instability onset is referred to as the occurrence of runaway deformation according to Lade and Pradel (1990), Vaid and Eliadorani (1998), Leong et al. (2000) and Chu et al. (2003). The instability is considered as liquefaction-reduced one when 5% double amplitude axial strain (peak-to-peak strain) or zero effective mean stress is reached, as suggested by Bray and Sancio (2006).

**Material**

The studied material is a loess taken from northern France located at 140 km North of Paris. Sampling was carried out according to the French Standard (AFNOR 1994): a trench of 1.5 x 9 m was opened, at a distance of 25 m from the railways; when reaching 1.2 m depth, a first sample block was prepared by cutting using a spade and then a knife; the block was put in a plastic box with the open side covered by a plastic film and then sealed using wax. The same operation was repeated for other depths. In total, four different depths (1.2, 2.2, 3.5 and 4.9 m) were considered. Figure 1 shows that the grain size distribution curves of four samples are comparable and typical of loess, with a slightly higher clay fraction (18 and 20%) in the 1.2 and 4.9 m samples respectively, as compared to 16% for the two other samples. The geotechnical properties of the samples are presented in Table 1. The liquid limit ($w_L$) varies from 26 to 30%; the plasticity index ($I_p$) varies from 6 to 9 %; the dry density ($\rho_d$) is rather low (1.39 - 1.55 Mg/m$^3$) that correspond to high porosities ($n$ = 0.43 – 0.49) or large void ratio ($e$ = 0.76 – 0.93); the carbonate content is quite high (% $CaCO_3$ = 5 - 15%). X-ray diffractometry showed that the clay fraction involves kaolinite, illite and interstratified illite-smectite. The suction measured using filter paper method (see detail in Delage et Cui 2000) is not high, ranging from 13.8 to 34.1 kPa, even though the natural degree of saturation (Sr) is quite low, comprised between 53 and 82% ( initial natural water content $w_{nat}$ = 16.6 – 23.7 %). Figure 2 shows the microstructure of the soil taken at 2.20 m depth, observed on Scanning Electronic Microscope (SEM). It can be observed that the soil is mainly composed of aggregates with the presence of large pores in a diameter bigger than 100 μm. The aggregates consist of silt grains bound together by fine materials. These materials must correspond to the clays and the carbonates, identified previously. Cui et al. (2004) made observations using mercury intrusion technique, showing that the collapse upon wetting correspond to the collapse of the larger pores; after collapse, a well graded pore size distribution corresponding to inter-grains pores was identified.

**Experimental methods**
Two types of cyclic triaxial tests were carried out, on initially saturated and unsaturated samples, respectively. Only non-reversal cyclic compression loading (without extension loading) was applied. Because of the limitation of the used equipment, the applied loading frequency was 0.05 Hz, much lower than that by the high speed train (20 Hz). Undrained cyclic shear tests on saturated specimens were conducted using the cyclic triaxial cell presented by Cui et al. (2007). Compared to a standard cell, the used cell allows monitoring the level change of water contained in the inner cell which surrounds the sample; the volume change of the sample can be then deduced from this water level change. Note that no suction probes are installed and therefore no suction measurement is allowed. The main reason is cyclic loading corresponds to quick loading and it would be too fast for any suction probes to follow the suction changes within the sample. Furthermore, it can also be suspected that the suction distribution would be not homogeneous during cyclic loading, tough at a loading frequency as low as 0.05 Hz. Therefore, for the first series of tests on initially saturated samples, the cell was used as a standard one with the monitoring of pore water pressure, whereas for the second series of tests on initially unsaturated samples, the volume change was monitored using the system of differential transducer. No suction was measured and the soils specimens were assumed to be at constant water state before reaching the saturated state.

*On initially saturated samples*
Cylindrical soil specimens of 70 mm in diameter and 140 mm high were prepared by cutting and trimming from soil blocks taken at four different depths (1.2m, 2.2m, 3.5m and 4.9m). After flushing with $CO_2$ gas for about 45 min, the specimens were saturated with de-aired water. 200 kPa back pressure was applied to obtain a satisfactory saturation with a Skempton's coefficient B higher than 0.99. The saturated specimen was then isotropically consolidated under an effective confining pressure of 25 kPa (the consolidation pressure $\sigma'_c = \sigma'_3$ = 25 kPa). The volume changes during the saturation and consolidation processes, which were estimated to be small, were not monitored. Force-controlled cyclic triaxial tests were then carried out under undrained condition by applying cyclic loads using a hydraulic piston. The cyclic deviator stress can therefore change with the section variation of soil sample. The effective confining pressure ($\sigma'_3$ = 25 kPa) was maintained constant during cyclic loading. The axial strain ($\varepsilon_1$), the pore water pressure change ($\Delta u$), the deviator stress (q = $\sigma_1$ - $\sigma_3$) and the effective mean pressure (p' = $\sigma'_1$/3 + 2$\sigma'_3$/3) were recorded using a data logging system. The deviator stress q was calculated by considering the imposed force F and a corrected specimen section $S_c$ as:

$$S_c = S_0 / (1 - \varepsilon_1)$$

where $S_0$ is the initial section.

Twelve cyclic tests were conducted, three for each depth (Table 2). For the soil of a given depth, an undrained triaxial test upon monotonic loading was first carried out, giving its peak deviator stress $q_{max}$. The 3 cyclic triaxial tests were performed under different deviator stresses $q_{cyc}$, within $q_{max}$. For instance, the peak deviator stress for the soil at 2.20 m is 14.9 kPa, the 3 cyclic tests were performed under 8.4, 11.7 and 14.9 kPa, respectively. An exception was made for the soil at 1.2 m: the first test was performed under 25 kPa deviator stress for 800 cycles and then under 50 kPa deviator stress.

*On initially unsaturated samples*

Only the soil at 2.20 m was studied because its highest collapse susceptibility identified by Cui et al. (2004) as well as the lowest cyclic resistance evidenced by the previous tests on initially saturated samples. Cylindrical specimens of 140 mm high and 70 mm in diameter were prepared by cutting and trimming from the soil blocks. After introducing it in a cylindrical latex membrane (70 mm in diameter), the soil was wetted by adding water drops using a pipette. The latex membrane was used to avoid the collapse of the soil specimen during wetting. When the desired water content was reached (calculated by considering the water content before water addition and the added water quantity), wetting was stopped and the soil was covered by a plastic film during 24 h for water content homogenization. The control of the sample dimension using an accurate clipper showed that the volume change due to wetting was negligible. This is consistent with its low plasticity nature. Three initial water contents $w_i$ were considered: 30, 32.2 and 33%. The corresponding degrees of saturation $S_{ri}$ are 85, 92 and 94%, respectively (Table 3).

After the installation of the soil specimen in the triaxial cell, a confining pressure of 25 kPa was applied under drained conditions. As in the previous case on initially saturated specimens, the volume change due to this pressure was estimated to be small and not monitored.

The first cyclic loadings that aimed at saturating the specimens by decreasing their volume at constant water content were done by trial and error: cyclic loading was applied by increasing progressively the cyclic deviator; it was stopped when the volume change of the soil specimen is equal to the initial air-pore volume in the soil. Assuming that the soil water content did not change during the cyclic loading, the suppression of air volume means that the soil became saturated. Note that this assumption may be relevant in the beginning of each test when the degree of saturation was relatively low and it can be no longer plausible when the degree of saturation became close to 100%. Thus the adopted method for the saturation control is not necessarily accurate.

After this phase, all the draining system (tubing and cell basis) was filled with de-aired water. Undrained cyclic test was then performed without application of back pressure. Nine tests, Test 13 to 21, were performed (Table 3); three tests for each water content.

**Tests results**

*On initially saturated samples*
Figure 3 presents the results from Test 6 performed on an initially saturated soil specimen. The figure is organized in four plots: (i) axial strain $\varepsilon_l$ versus cycle number $N$; (ii) deviator stress $q$ versus axial strain $\varepsilon_l$ ; (iii) pore water pressure change $\varDelta u$ versus axial strain $\varepsilon_l$ ; and (iv) deviator stress $q$ versus effective mean net stress $p'$. It can be observed in the $\varepsilon_l$ - $N$ plot that $\varepsilon_l$ increases slightly for the first 3 cycles and then abruptly afterwards: the mean slope of the curve presents a significant increase after the first 3 cycles. The double amplitude strain (peak-to-peak strain) is about 3% at the fourth cycle and more than 7% at the fifth cycle. The final strain reached 19% in less than 7 cycles. Interestingly, $\varDelta u$ increases drastically for the first 3 cycles, to reach a value higher than 17 kPa at the end of the third cycle. The final value of $\varDelta u$ is about 22.7 kPa, close to the confining pressure of 25 kPa. Examination of $q$ - $\varepsilon_l$ plot shows that q started to decrease significantly when $\varepsilon_l$ started to develop at a higher rate from the third cycle; this is obviously the consequence of the specimen's section correction. In the $p'$- $q$ plot, in addition to q decrease, a progressive decrease of $p'$ was observed, as a result of $\varDelta u$ increase. As defined by Lade and Pradel (1990), Vaid and Eliadorani (1998), Leong et al. (2000) and Chu et al. (2003), the point corresponding to the abrupt change in $\varepsilon_l$ rate can be referred to as the instability state of the soil, it can be concluded that for Test 6, the instability was reached at about 2% axial strain after 3 cycles.

In total, twelve tests were performed and instability was observed in 10 tests: for the 3 tests on the soil at 3.50 m, instability was reached only under $q_{cyc}$ = 25 kPa. All the results are gathered in Table 2 with the values of deviator stress $q_{ins}$ and cycle number $N_{ins}$ in instability state, as well as the cyclic resistance defined as $\frac{\tau_{cyc}}{\sigma'_c} = \frac{q_{ins}}{2\sigma'_c}$. Comparison between Test 1 (800 cycles under $q_{cyc}$ = 25 kPa followed by cyclic loading under $q_{cyc}$ = 50 kPa) and Test 2 (cyclic loading under $q_{cyc}$ = 50 kPa only) showed that a previous cyclic loading that did not cause the soil instability led to a smaller value of $N_{ins}$ under higher deviator stress. Indeed, Test 1 showed instability onset at $N_{ins}$ = 15 against $N_{ins}$ = 18 from Test 2. For other tests, $N_{ins}$ ranged from 1 to 145. The cyclic resistance, $\tau_{cyc}/\sigma'_c$, varies from 0.17 to 1.40.

The results of cyclic resistance from all the tests are plotted versus $N_{ins}$ in Figure 4. Three regression lines are plotted for the three depths: 1.2, 2.2 and 4.9 m. It is observed that for all the depths, the lower the value of $\tau_{cyc}/\sigma'_c$ the bigger the number of cycles $N_{ins.}$ Comparison between the four depths showed that the soil at 1.20 m is the most resistant and the soil at 2.20 m is the less stable.

*On initially unsaturated samples*
Figure 5 presents the results from Test 20 (w = 33%) during the first cyclic loading stage. As the q – N plot (Figure 5a) indicates, the samples was loaded under 4.7 kPa for 30 cycles, 11.7 kPa for 20 cycles, 21.5 kPa for 125 cycles, 25.5 kPa for 27 cycles and 41 kPa for 8 cycles. The $\varepsilon_1$-N plot shows that this cyclic loading induced an axial strain $\varepsilon_1$ up to 2.3% (Figure 5b) and the volume decrease reached about 11.4 cm$^3$ which corresponds to a volumetric strain of 4.5% and a void ratio change from $e_i$ = 0.93 to $e_f$ = 0.86. The total volume changes and the final void ratios for all the tests after the first cyclic loading stage are presented in Table 3. It appears that for a given soil with a same loading procedure, the final volume change was not the same. This can be mainly explained by the difficulty of choosing accurate deviator stress during the first loading stage. Indeed, when approaching the target volume change, it was difficult to precisely define the deviator value that allowed obtaining the target volume change in a reasonable duration.

After this first stage of cyclic loading, the cell basis and the drainage tubing were saturated and the second cyclic loading under undrained conditions started. Figure 6 presents the results of Test 20. Cyclic deviator loading was performed with $q_{cyc}$ = 27 kPa. The $\varepsilon_1$-N plot (Figure 6a) shows that $\varepsilon_1$ varied slightly and almost linearly with cycle number N at $\varepsilon_1$ < 2.5%. The variation started to speed up afterward, reaching 37% at N = 47. The threshold value of 2.5% for low strain rate is similar to that in the case of initially saturated sample (about 2%). This led us to adopt the same instability criterion in terms of changes in $\varepsilon_1$ rate: for Test 20, the soil instability occurred at $\varepsilon_1$ < 2.5%, $N_{ins}$ = 12. Note that as opposed to Test 6, an accumulation of axial strain without amplification of amplitude was observed in test 20; the amplitude remained small even at $\varepsilon_1$ > 35%. The variation of $\Delta u$ showed in Figure 6b presents also two distinguished rates: a quick change followed by a slower one. Interestingly, the transition point is at about 2.5% axial strain. Figure 6c shows that the deviator stress q decreases in agreement with the axial strain accumulation and Figure 6d shows p' decrease as the consequence of $\Delta u$ increase.

The values of maximum cyclic deviator $q_{cyc}$, deviator at the onset of instability $q_{ins}$, cycle number at the onset of in stability $N_{ins}$, and cyclic resistance $\dfrac{\tau_{cyc}}{\sigma_c^{'}} = \dfrac{q_{ins}}{2\sigma_c^{'}}$, for all the 9 tests are gathered in Table 3. It can be observed that at low initial water content and low deviator stress, a very big number of cycles was needed to reach the instability state. For instance, $N_{ins}$ = 1400 under $q_{cyc}$ = 45 kPa at $w_i$ = 30%; $N_{ins}$ = 915 under $q_{cyc}$ = 40 kPa at $w_i$ = 32.2%.

Figure 7 presents the cyclic resistance versus cycle number at instability onset for the soil at 2.20 m at four initial water contents including that in saturated state (34%). Regression lines were added to represent the cyclic resistance curves. It appears clearly that the cyclic resistance strongly depends on the initial water content: the lower the value of $w_i$ the higher the cyclic resistance.

**Discussion**

In the present work, the instability onset was identified from $\varepsilon_1$-N plot where $\varepsilon_1$ rate meets rapid change (Figures 3 and 6). This rapid change occurs in general at 1.5 to 3% of axial strain. In terms of pore water changes, this point also corresponds to the transition from quick to lower variation. Note that the observed instability is consistent with the instability state identified mostly on remold saturated sandy soils (Lade and Pradel 1990, Vaid and Eliadorani 1998, Leong et al. 2000, Chu et al. 2003). This instability is however not necessarily corresponding to the liquefaction state according to the two widely admitted liquefaction criteria: one is based on the amplitude of axial strain (5% in general, see Wang et al. 2006; Ueng et al. 2004; Hyde et al. 2006; among others) and another on zero effective mean stress.

Bray and Sancio (2006) worked on undisturbed fine-grained soils and showed that liquefaction occurs when one of the following conditions is verified: (i) 3% single amplitude axial strain is reached; (ii) 5% double amplitude axial strain (peak-to-peak strain) is reached;

(iii) the excess pore water pressure reached the confining pressure. Furthermore, by comparing the results obtained with two frequencies, 0.005 Hz and 1 Hz, Bray and Sancio (2006) noted that the strain/deformation-based criterion for the onset of liquefaction is not influenced by the frequency. Table 4 summarizes different data in i) the instability state according to the defined criterion (cycle number N and axial strain $\varepsilon_1$), ii) the liquefaction state according to the 5% peak-to-peak axial strain criterion (cycle number N and pore water pressure $\Delta u$) and iii) the liquefaction state according to the pore water pressure-based criterion (cycle number N, pore water pressure $\Delta u$ and axial strain $\varepsilon_1$). Note that only two tests (Test 10 and 12) showed an absolute zero effective mean stress condition with pore water pressure equal to the confining pressure of 24.9 kPa. In the following analysis, liquefaction is regarded as reached when water pressure is higher than 22 kPa (against 25 kPa effective confining pressure). It can be observed that
- only Test 7 and 8 on the soil from 3.5 m depth did not show instability;
- 9 tests (Test 3, 5, 6, 14, 15, 16, 17, 18, 21) showed liquefaction according the strain/deformation-based criterion;
- 8 tests (Test 1, 4, 5, 6, 9, 10, 11, and 12) showed liquefaction according to the pore water pressure-based criterion.

In addition, comparison between the number of cycles to reach instability and that to reach liquefaction shows that there is a general good agreement between the instability criterion and the strain/deformation-based criterion, and on the contrary, the number of cycles for pore water pressure based criterion is generally much larger. This means that the instability or the liquefaction according to the strain/deformation-based criterion occurs much before the liquefaction according to the pore water pressure-based criterion. This phenomenon is probably related to the particular microstructure of the naturally cemented loess. As Fig. 1 showed, the studied loess is constituted of aggregates with large inter-aggregate pores and small intra-aggregate pores. The presence of bonding materials as clays and carbonates contributes to its particular porous microstructure. Upon cyclic loading, this microstructure was progressively destabilized and it finally collapsed. This collapse might correspond to a large axial strain as identified according to the instability criterion and strain/deformation-based liquefaction criterion. However, although the significant pore water pressure build-up during this stage, it was not high enough to balance the total mean stress. Further cyclic loading was needed to reach the zero effective mean stress condition. Many authors also showed that cementation hinders the early build-up of pore water pressure (Mitchell and Solymar 1984; Clough et al. 1989; Leon et al. 2006, Yun and Santamarina 2005). For soil constituted of silt and clay, the criteria used to define the stage of initial liquefaction for sands may not be applicable, because of the difference in pore water pressure variation generation and deformation relationship as compared with those of sand (Guo and Prakash 1989; Ueng et al. 2004; Bray and Sancio 2006).

Several works in the literature show that the soil's cyclic resistance increases with its relative density (Seed and Idriss 1971; Seed 1979; Castro and Poulos 1977; Vaid and Chern 1985; Kramer and Seed 1988; Alarcon-Guzman et al. 1988), fines content (Ishihara 1985; Kuerbis et al. 1988; Erten and Maher 1995; Singh 1994 and Singh 1996), plasticity index (Bray and Sancio 2006), consolidation pressure (Casagrande 1976; Tokimatsu and Hosaka 1986; Yasuda et al. 1997), aging effect (Mitchell and Solymar 1984; Skempton 1986; Kulhawy and Mayne 1990) and cementation (Acar and EL-Tahir 1986; Lade and Overton 1989; Fernandez and Santamarina 2001; Yun and Santamarina 2005; Mohsin and Airey 2005). Examination of the cyclic resistance curves obtained for the four depths (Figure 4) shows that the soil at 2.2 m has the lowest resistance and the soil at 1.2 m depth has the highest resistance. As showed, the soil at 2.2 m has the highest porosity, n = 0.49, and the soil at 1.2 m has the highest fines content (20%) and largest plasticity index. The effects of porosity, fines content, and plasticity index observed in the present work are then in agreement with the results obtained in the literature.

Figure 7 evidences that the cyclic resistance was strongly affected by the initial water content. In fact, for the tests on initially saturated samples, the samples were in near intact state, and therefore with low density (Table 1). On the contrary, for the tests on initially unsaturated samples, the latter were subjected to a "saturation" process upon cyclic loading under near constant water content condition. This process induced volume decrease and thus a densification of the samples. As mentioned above, this densification might increase the soil cyclic resistance. In addition, the lower the initial water content or the lower the initial degree of saturation, the more the densification, and thus the higher the cyclic resistance. Therefore the effect of initial water content should be rather regarded as the effect of initial density.

**Conclusion**

Triaxial cyclic loading tests were performed on loess taken near a high-speed train railway in Northern France. Two types of tests were conducted, one on initially saturated samples and another on initially unsaturated samples. For the former, soil samples were saturated under 200 kPa back-pressure and then loaded under different non-reversal cyclic deviator stresses that were estimated based on the results from monotonic shear tests. For the latter, soil samples were first subjected to cyclic loading under near constant water content condition; when their volume decrease led to the saturation state, they were loaded under undrained condition as the initially saturated samples. The instability onset was identified from $\varepsilon_1$-N plot where $\varepsilon_1$ rate meets rapid change. This rapid change occurs in general at 1.5 to 3% axial strain. This instability is similar to the instability state identified mostly on remold saturated sandy soils (Lade and Pradel 1990, Vaid and Eliadorani 1998, Leong et al. 2000, Chu et al. 2003). According to this instability-based criterion, the soil having the highest porosity and lowest fines contents (at 2.2 m depth) has the lowest cyclic resistance, and the soil having the highest fines content and largest plasticity index (at 1.2 m depth) is the most resistant. The effect of initial water content was evidenced: the lower the initial water content the higher the cyclic resistance. This effect should be regarded as the effect of initial densification due to the "saturation" stage by loading under near constant water content condition.

Compared to the common strain/deformation-based liquefaction criterion, the instability-based criterion has been found to be more general since it covers not only the case of increasing axial strain amplitude with cycling but also the case of axial strain accumulation without clear amplitude increase (case of fatigue). It has been observed that in case of liquefaction, there was a good agreement between the two criteria. Nevertheless, no consistency was observed between the strain/deformation-based liquefaction criterion and the pore water pressure-based liquefaction criterion. This is probably related to the particular microstructure of naturally cemented soils as the studied loess: the strain/deformation-based criterion would correspond to the initial microstructure collapse whereas more cycles were needed to verify the pore water pressure-based criterion.

From a practical point of view, when this instability occurs, significant volume change can take place after dissipation of the generated water pressure. This volume change seems to be large enough to explain the soil instability observed in the field. It should be however recalled that the test conditions are not exactly the same as what happens in the field: the applied confining pressure of 25 kPa was estimated based on the results of numerical analysis; the applied cyclic deviator stresses were higher than that recorded in field in order to drive the soil beyond the instability state. Further studies are therefore necessary under a condition close to the real one.

**Acknowledgements**
The authors address their thanks and gratitude to French Railway Company (SNCF) and RFF (Réseau Ferré de France) for their financial support.

Table 1. Geotechnical properties of the studied loess

| Depth (m) | %<2μm | $w_L$ (%) | $I_p$ | $\rho_d$ (Mg/m$^3$) | n | $w_{nat}$ (%) | e | $S_r$ (%) | CaCO$_3$ (%) | Suction (kPa) |
|---|---|---|---|---|---|---|---|---|---|---|
| 1.2 | 20 | 30 | 9 | 1.52 | 0.44 | 18.9 | 0.76 | 66 | 5 | 20 |
| 2.2 | 16 | 28 | 9 | 1.39 | 0.49 | 18.1 | 0.93 | 53 | 6 | 34 |
| 3.5 | 16 | 26 | 6 | 1.54 | 0.43 | 16.6 | 0.82 | 55 | 15 | 27 |
| 4.9 | 18 | 30 | 9 | 1.55 | 0.43 | 23.7 | 0.78 | 82 | 9 | 14 |

Table 2: Testing program for initially saturated samples

| Test No | depth (m) | e | $q_{max}$ (kPa) | $q_{cyc}$ (kPa) | $q_{ins}$ (kPa) | $N_{ins}$ (cycles) | $\frac{\tau_{cyc}}{\sigma_c'} = \frac{q_{ins}}{2\sigma_c'}$ |
|---|---|---|---|---|---|---|---|
| 1 | 1.20 | 0.76 | 71.0 | 50.0 | 48.0 | 15 | 0.96 |
| 2 |  |  |  | 50.0 | 48.0 | 18 | 0.96 |
| 3 |  |  |  | 70.0 | 50.0 | 1 | 1.40 |
| 4 | 2.20 | 0.93 | 14.9 | 8.4 | 8.4 | 106 | 0.17 |
| 5 |  |  |  | 11.7 | 11.7 | 28 | 0.23 |
| 6 |  |  |  | 14.9 | 14.7 | 3 | 0.29 |
| 7 | 3.50 | 0.82 | 22.9 | 15.0 | - | - | - |
| 8 |  |  |  | 22.0 | - | - | - |
| 9 |  |  |  | 25.0 | 22.0 | 30 | 0.44 |
| 10 | 4.90 | 0.78 | 49.7 | 15.0 | 30.0 | 145 | 0.60 |
| 11 |  |  |  | 30.0 | 35.0 | 35 | 0.70 |
| 12 |  |  |  | 48.0 | 47.0 | 25 | 0.94 |

- Instability did not occur.

Table 3: Testing program for initially unsaturated samples

| Test N$^0$ | $w_i$ (%) | $S_r$ (%) | $\Delta V/V_0$ | $e_f$ | $q_{cyc}$ (kPa) | $q_{ins}$ (kPa) | $N_{ins}$ | $\frac{\tau_{cyc}}{\sigma_c'} = \frac{q_{ins}}{2\sigma_c'}$ |
|---|---|---|---|---|---|---|---|---|
| 13 | 30.0 | 85 | 0.062 | 0.81 | 44 | 42 | 1400 | 0.84 |
| 14 |  |  | 0.090 | 0.75 | 78 | 77 | 115 | 1.54 |
| 15 |  |  | 0.057 | 0.82 | 95 | 95 | 70 | 1.90 |
| 16 | 32.2 | 92 | 0.034 | 0.86 | 42 | 40 | 915 | 0.80 |
| 17 |  |  | 0.044 | 0.84 | 81 | 76 | 30 | 1.52 |
| 18 |  |  | 0.050 | 0.83 | 90 | 83 | 10 | 1.66 |
| 19 | 33.0 | 94 | 0.021 | 0.89 | 20 | 20 | 100 | 0.40 |
| 20 |  |  | 0.045 | 0.86 | 27 | 27 | 12 | 0.54 |
| 21 |  |  | 0.035 | 0.84 | 40 | 40 | 2 | 0.80 |

Table 4: Instability or liquefaction identified using different criteria

| Conditions | | Defined instability criterion | | Strain/deformation-based liquefaction criterion | | Pore water pressure-based liquefaction criterion | | |
|---|---|---|---|---|---|---|---|---|
| **Saturated soil** | | N (cycles) | $\varepsilon_1$ (%) | $N_{DA}$ (cycles) | $\Delta u$ (kPa) | $\Delta u_f$ (kPa) | N (cycles) | $\varepsilon_1$ (%) |
| Z(m) | Test | | | | | | | |
| 1.20 | 1 | 15 | 2 | - | - | 24.4 | 105 | 21 |
| | 2 | 18 | 2 | - | - | 19 | 110 | 21 |
| | 3 | 1 | 0.25 | 4 | 18 | - | - | - |
| 2.20 | 4 | 106 | 2 | - | - | 23.3 | 155 | 19 |
| | 5 | 28 | 0.8 | 37 | 23 | 23.4 | 46 | 19 |
| | 6 | 3 | 2 | 5 | 22 | 22.7 | 7 | 19 |
| 3.50 | 7 | - | - | - | - | - | - | - |
| | 8 | - | - | - | - | - | - | - |
| | 9 | 50 | 2 | - | - | 24.5 | 393 | 19 |
| 4.90 | 10 | 145 | 1.45 | - | - | 24.9 | 185 | 13 |
| | 11 | 35 | 2 | - | - | 24.4 | 200 | 20 |
| | 12 | 25 | 2 | - | - | 24.9 | 60 | 20 |
| **Unsaturated soil (2.20m)** | | N (cycles) | $\varepsilon_1$ (%) | $N_{DA}$ (cycles) | $\Delta u$ (kPa) | $\Delta u_f$ (kPa) | N (cycles) | $\varepsilon_1$ (%) |
| $w_i$(%) | Test | | | | | | | |
| 30 | 13 | 1400 | * | - | - | * | * | - |
| | 14 | 115 | 1.3 | 130 | 10 | 20 | 140 | 29 |
| | 15 | 70 | 1.5 | 80 | 11 | 19 | 90 | 30 |
| 32.2 | 16 | 915 | 1.3 | 950 | 16 | 20 | 1005 | 20 |
| | 17 | 30 | 1 | 55 | 12 | - | - | - |
| | 18 | 10 | 1.25 | 31 | 10 | - | - | - |
| 33 | 19 | 100 | 0.5 | - | - | - | - | - |
| | 20 | 12 | 2.5 | - | - | 19.9 | 50 | 37 |
| | 21 | 2 | 2.5 | 2 | 9 | - | - | - |

**- *Instability or liquefaction not observed***
***  Data not available***

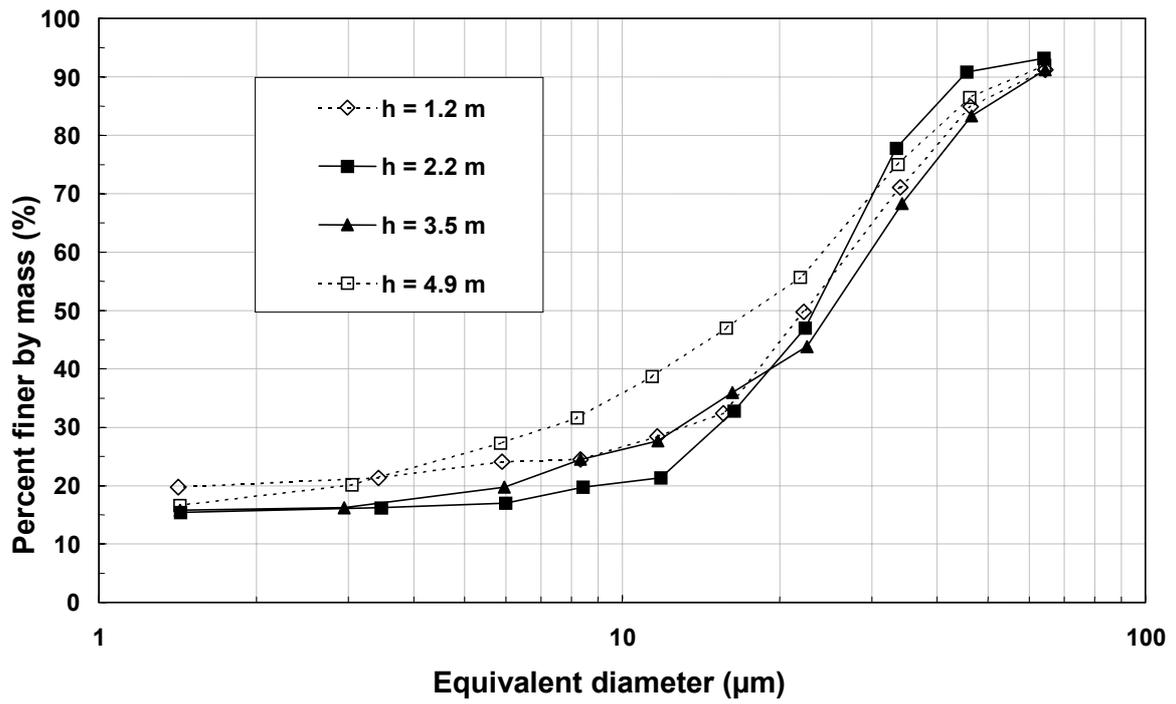

Figure 1. Grain size distribution curves of the four samples

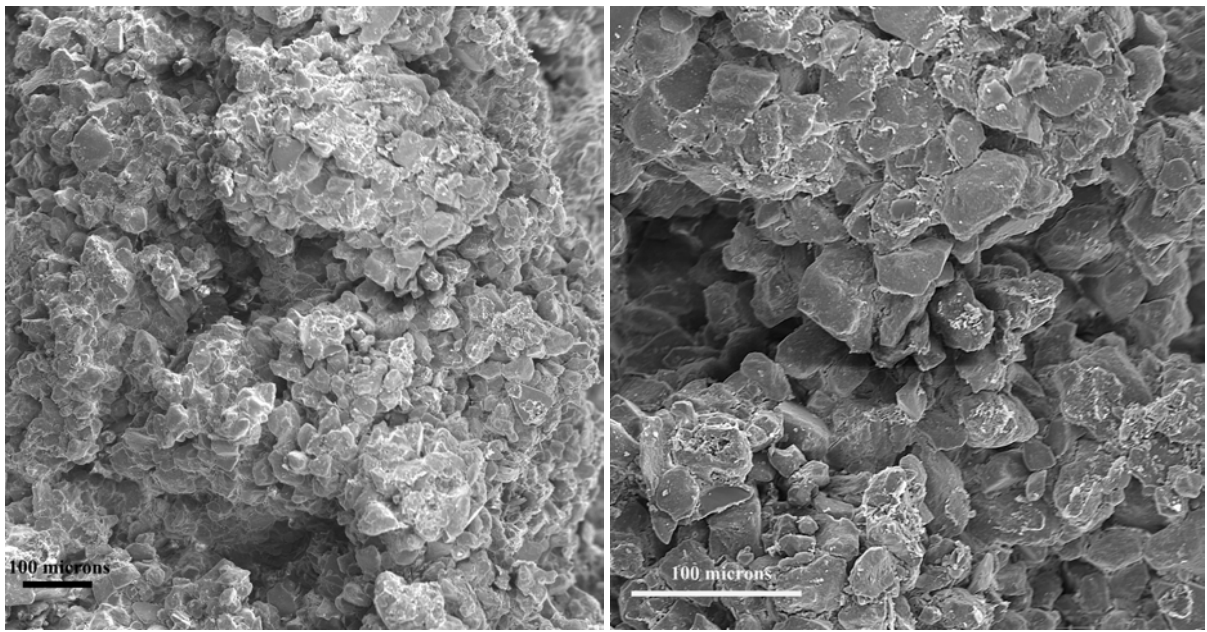

Figure 2.   Observation of loess taken at 2.20 m depth on Scanning Electron Microscope (SEM)

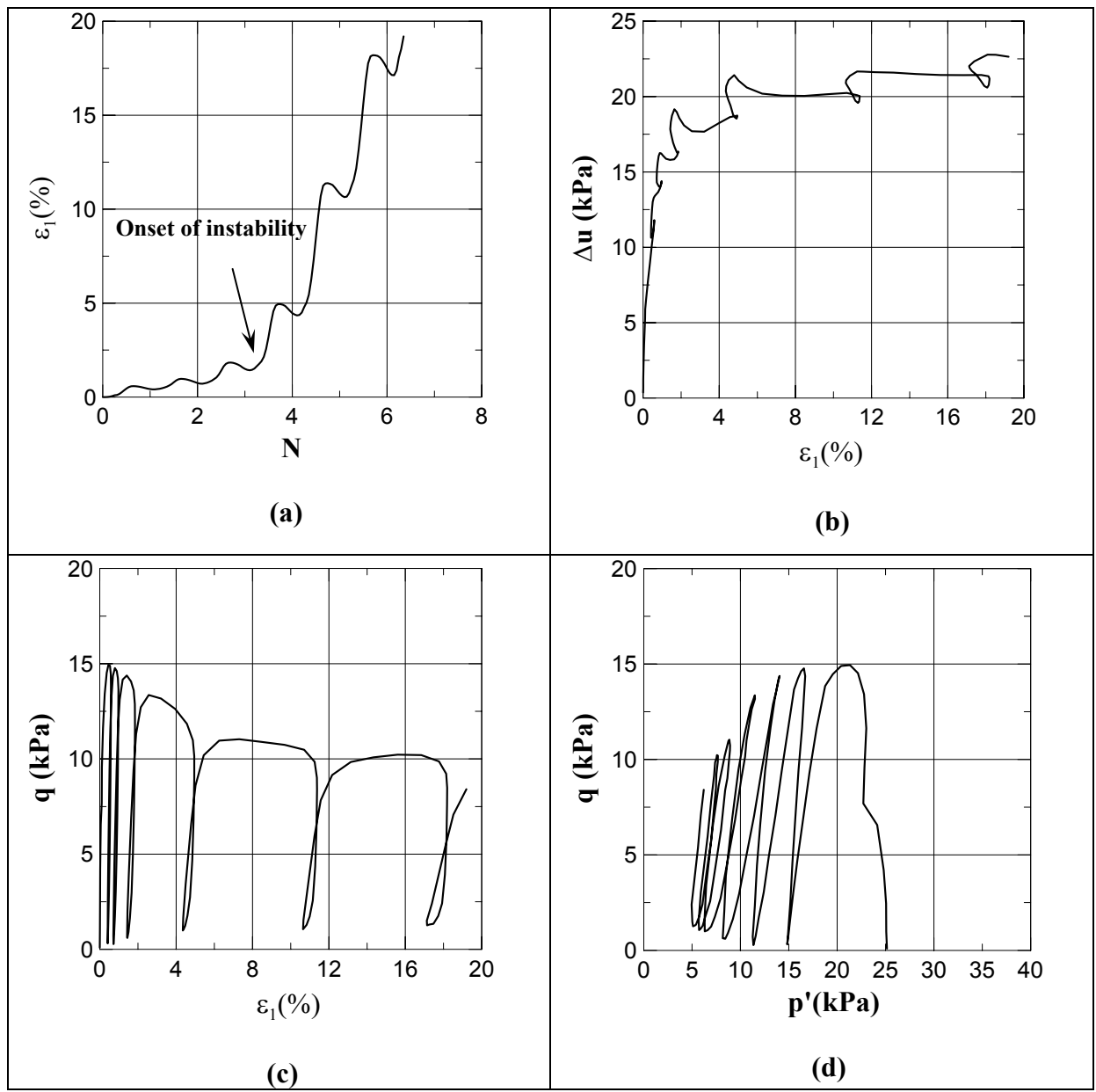

**Figure 3.** Results from cyclic triaxial tests conducted on loess taken at 2.2 m (a) $\varepsilon_1 - N$ (b) $\Delta u - \varepsilon_1$ (c) $q - \varepsilon_1$ (d) $q - p'$

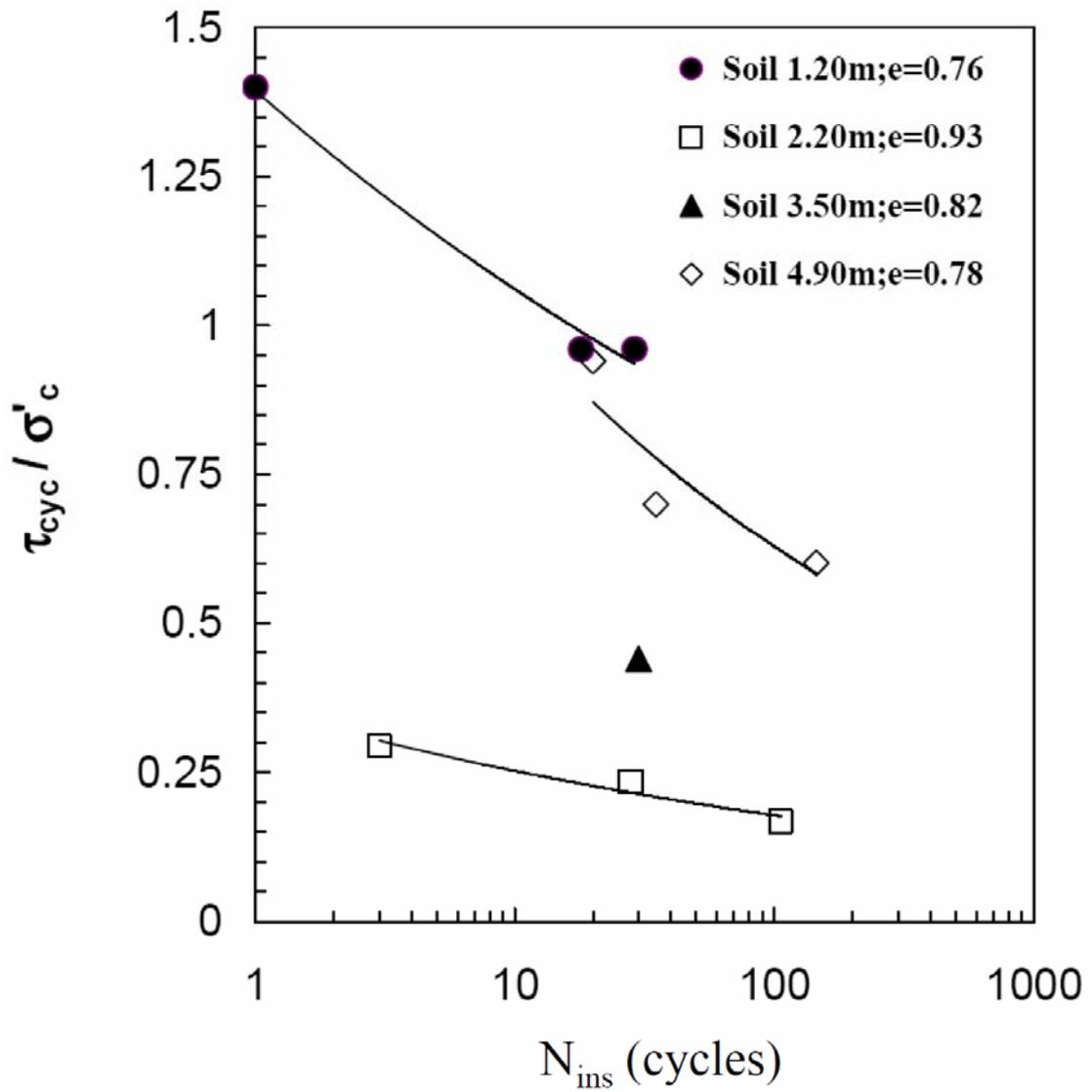

**Figure 4.** Cyclic resistance curves obtained from cyclic triaxial tests conducted on saturated loess samples taken at different depths

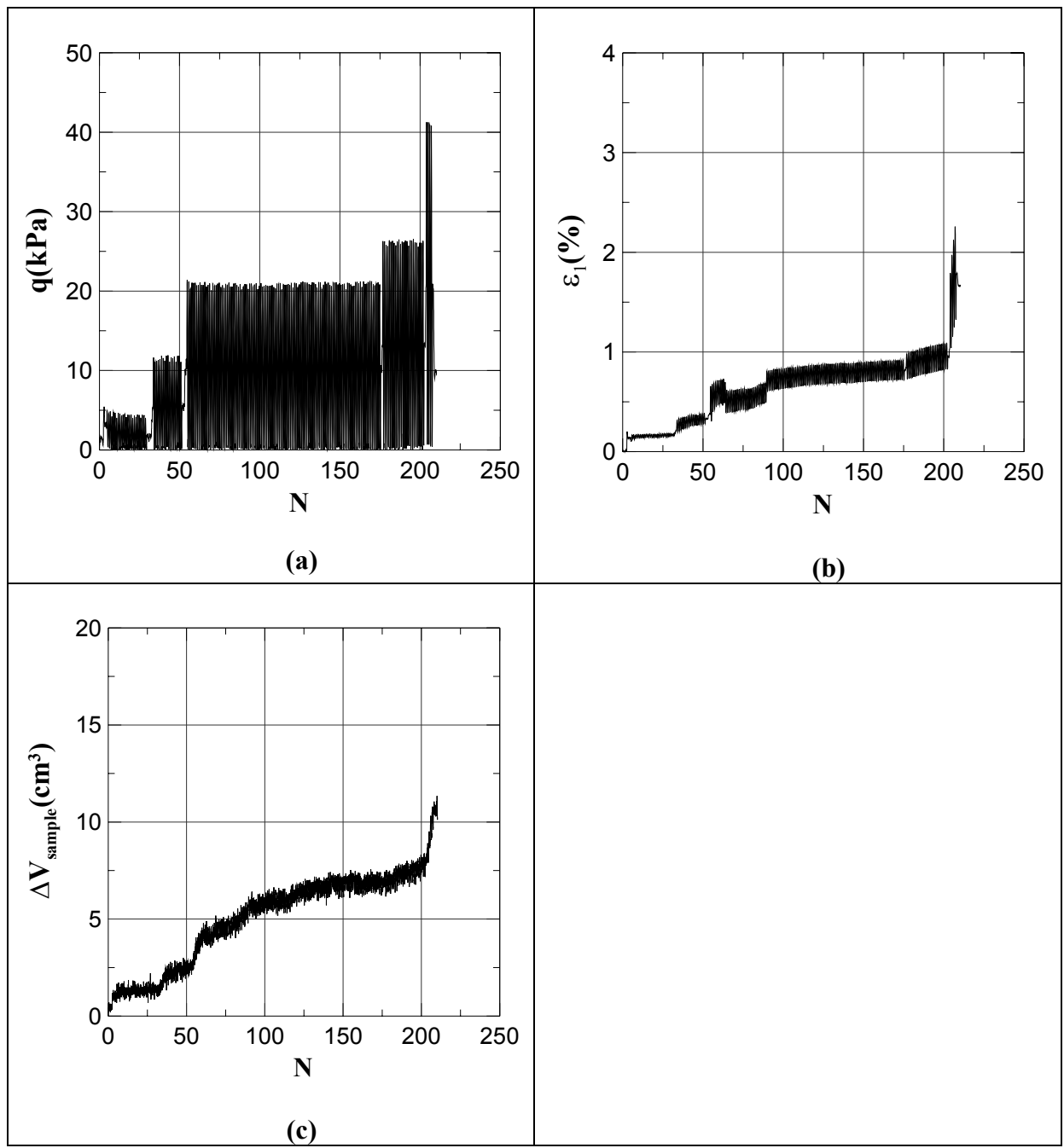

**Figure 5.** Results from cyclic triaxial tests conducted on loess taken at 2.2 m and humidified to w=33 % (a) q–N (b) $\varepsilon_1$–N (c) $\Delta$V–N.

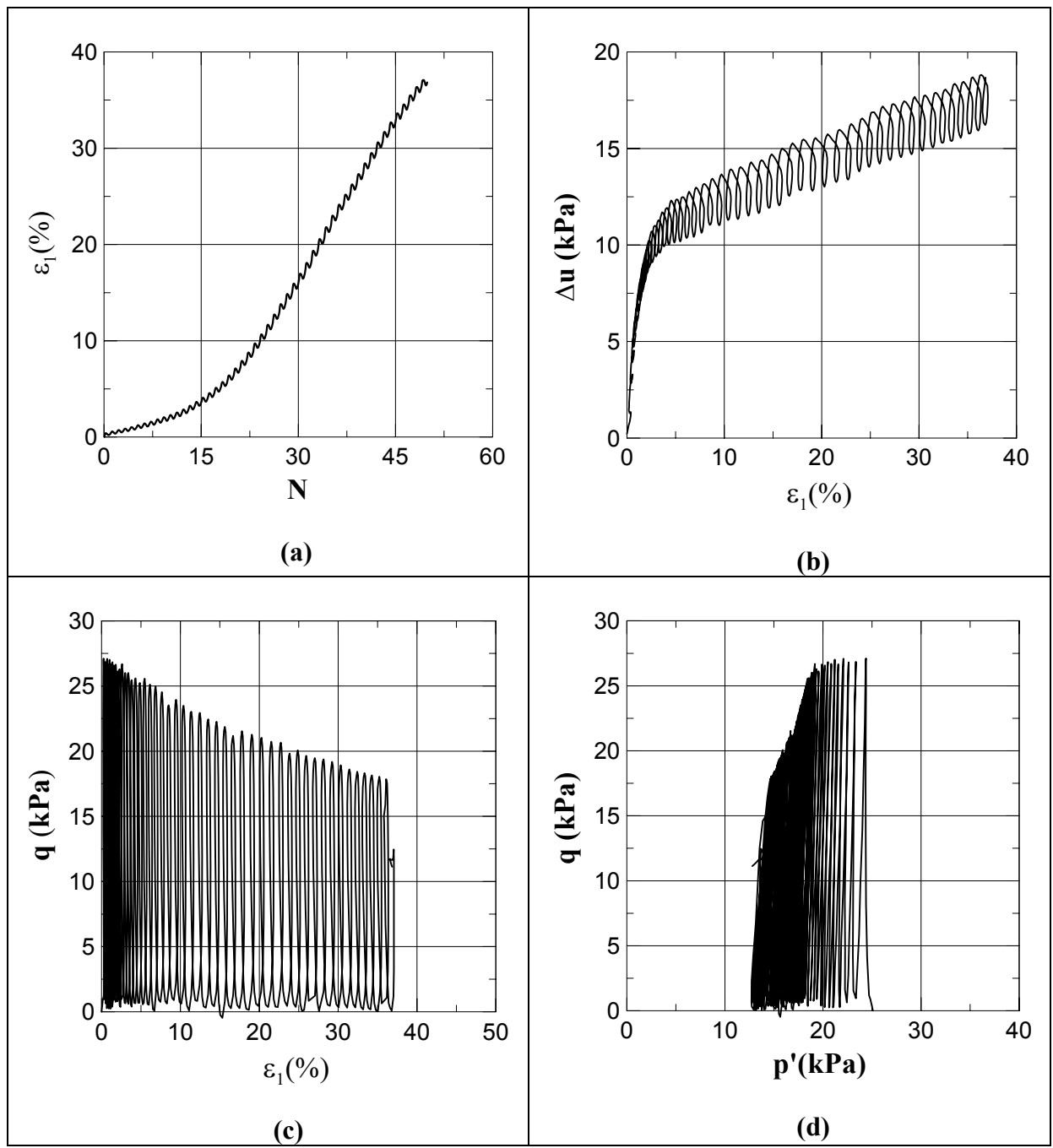

**Figure 6.** Results from cyclic triaxial tests conducted on loess taken at 2.2 m wetted at w=33% then loaded to "saturation" state. (a) $\varepsilon_1$–N (b) $\Delta u$–$\varepsilon_1$ (c) q-$\varepsilon_1$ (d) q-p′

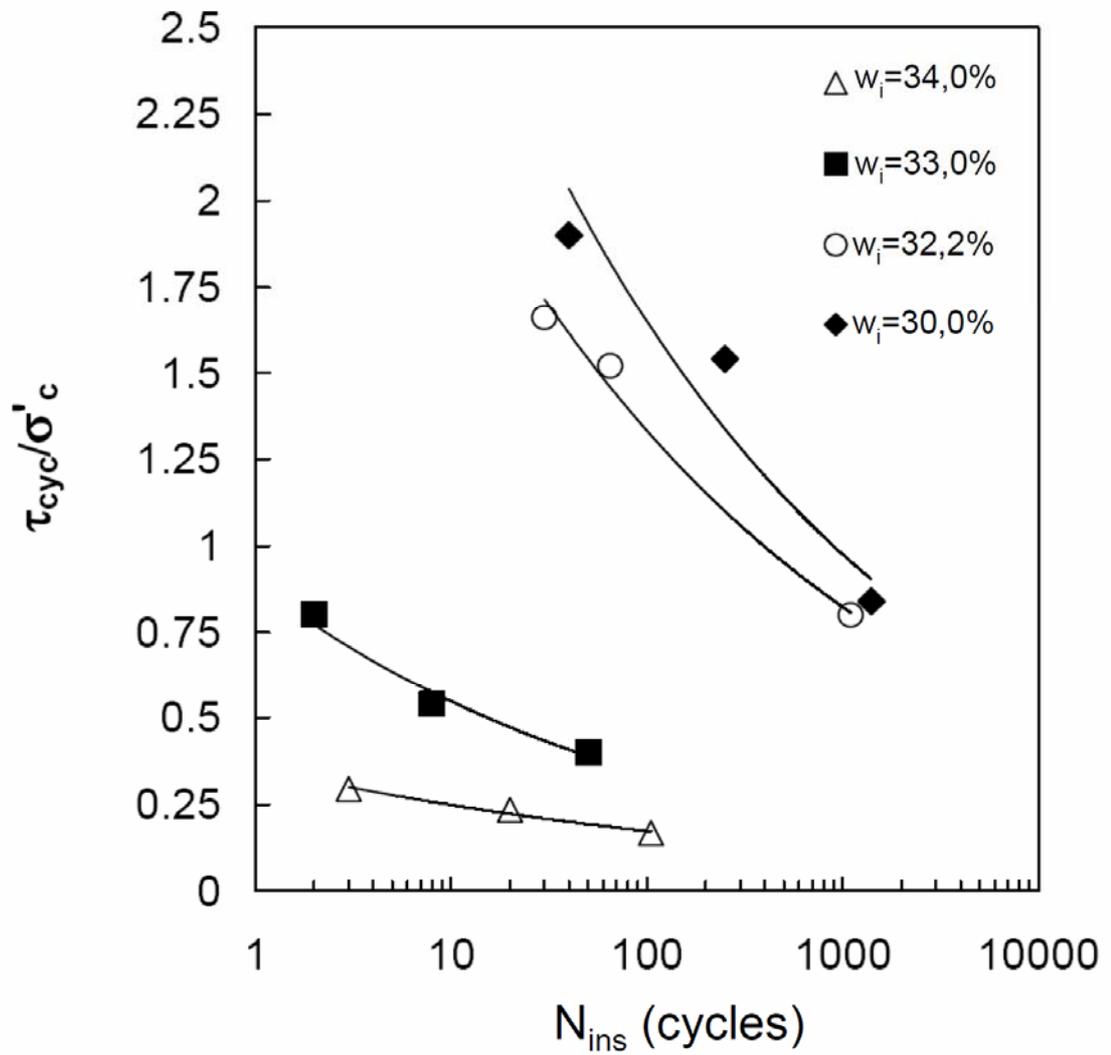

**Figure 7.** Cyclic resistance curves obtained from cyclic triaxial tests conducted on initially nearly saturated loess taken at 2.2 m with different initial hydrous states